\documentclass[11pt]{article}
\usepackage{cite}
\usepackage{mathrsfs}
\usepackage{amssymb,amsfonts,amsmath}
\usepackage{enumerate}
\usepackage{indentfirst}
\usepackage{latexsym,bm}
\usepackage[noend]{algpseudocode}
\usepackage{algorithmicx,algorithm}
\usepackage{algorithm}
\usepackage{makeidx}
\usepackage{fancybox}
\usepackage{color}
\usepackage{multicol}
\usepackage[latin1]{inputenc}
\usepackage{graphicx}
\usepackage{pstricks,pst-node,pst-text,pst-3d}
\usepackage{tikz}
\newtheorem{thm}{Theorem}[section]
\newtheorem{cor}[thm]{Corollary}
\newtheorem{lem}[thm]{Lemma}

\newtheorem{obs}[thm]{Observation}

\newenvironment{pf}{{\noindent \it \bf Proof:}}{{\hfill$\Box$}\\}

\newcommand{\2}{\vspace{0.2cm}}

\begin{document}

\title{\bf Strong Subgraph $k$-connectivity}

\author{Yuefang Sun$^{1}$, Gregory Gutin$^{2}$, Anders Yeo$^{3,4}$ and 
Xiaoyan Zhang$^{5}$\thanks{Corresponding author.} \\ \\
$^{1}$ Department of Mathematics,
Shaoxing University\\
Zhejiang 312000, P. R. China, yuefangsun2013@163.com\\
$^{2}$ Department of Computer Science\\
Royal Holloway, University of London\\
Egham, Surrey, TW20 0EX, UK, g.gutin@rhul.ac.uk\\
${}^3$Department of Mathematics and Computer Science \\
University of Southern Denmark \\
Campusvej 55, 5230 Odense M, Denmark, andersyeo@gmail.com \\
${}^4$Department of Pure and Applied Mathematics\\
University of Johannesburg \\
Auckland Park, 2006 South Africa \\
$^{5}$ School of Mathematical Science \& Institute of Mathematics\\
Nanjing Normal University\\
Jiangsu 210023, P. R. China, zhangxiaoyan@njnu.edu.cn\\
}

\date{}
\maketitle

\begin{abstract}
Generalized connectivity introduced by Hager (1985) has been studied extensively in undirected graphs and become an established area in undirected graph theory. For connectivity problems, directed graphs can be considered as generalizations of undirected graphs. In this paper, we introduce a natural extension of generalized $k$-connectivity of undirected graphs to directed graphs (we call it strong subgraph $k$-connectivity) by replacing connectivity with strong connectivity. We prove NP-completeness results and the existence of polynomial algorithms. We show that strong subgraph $k$--connectivity is, in a sense, harder to compute than generalized $k$-connectivity. However, strong subgraph $k$-connectivity can be computed in polynomial time for semicomplete digraphs and symmetric digraphs. We also provide sharp bounds on strong subgraph $k$-connectivity and pose some open questions.
\vspace{0.3cm}
\end{abstract}


\section{Introduction}\label{sec:intro}

Connectivity is one of
the most basic concepts in graph theory\footnote{We refer the readers to \cite{Bang-Jensen-Gutin, Bondy} for graph theoretical
notation and terminology not given here.}, both in combinatorial
and algorithmic senses. The classical connectivity has
two equivalent definitions. The connectivity of an undirected graph $G$, written
$\kappa(G)$, is the minimum size of a vertex set $S\subseteq V(G)$
such that $G-S$ is disconnected or has only one vertex. This
definition is called the {\em cut-version} definition.
The well-known theorem of Menger provides an equivalent
definition, which can be called the {\em path-version} definition.
For two distinct vertices $x$ and $y$ in $G$,
the local connectivity $\kappa_{\{x,y\}}(G)$ is the maximum number
of internally disjoint paths connecting $x$ and $y$. Then $\kappa(G)
= \min\{\kappa_{\{x,y\}}(G)\mid x,y \in V(G), x\neq y\}$ is defined
to be the connectivity of $G$.

The generalized $k$-connectivity $\kappa_k(G)$ of a graph $G$ which
was introduced by Hager \cite{Hager} in 1985, is a natural
generalization of the path-version definition of the connectivity.
For a graph $G=(V,E)$ and a set $S\subseteq V$ of at least two
vertices, an {\em $S$-Steiner tree} or, simply, an {\em $S$-tree}
 is a subgraph
$T$ of $G$ which is a tree with $S\subseteq V(T)$. Two $S$-trees
$T_1$ and $T_2$ are said to be {\em internally disjoint} if
$E(T_1)\cap E(T_2)=\emptyset$ and $V(T_1)\cap V(T_2)=S$. The {\em
generalized local connectivity} $\kappa_S(G)$ is the maximum number
of internally disjoint $S$-trees in $G$. For an integer $k$ with
$2\leq k\leq n$, the {\em generalized $k$-connectivity} is defined
as
$$\kappa_k(G)=\min\{\kappa_S(G)\mid S\subseteq V(G), |S|=k\}.$$
Observe that
$\kappa_2(G)=\kappa(G)$.
If $G$ is disconnected and vertices of $S$ are placed in different connectivity components, we have  $\kappa_S(G)=0$.
Thus, $\kappa_k(G)=0$ for a disconnected graph $G$.

Both extremes for $k$ in $\kappa_k(G)$ relate to fundamental theorems in
combinatorics. For $k=2$,  internally disjoint $S$-trees are
internally disjoint paths between the two vertices, and so the
parameter is relevant to the well-known Menger theorem. For $k=n$,
internally disjoint $S$-trees are
edge-disjoint spanning trees of the graph, and so this
parameter is relevant to the spanning tree packing problem
\cite{Ozeki-Yamashita, Palmer} and the classical Nash-Williams-Tutte
theorem \cite{Nash-Williams, Tutte}. Generalized
connectivity of graphs has become an established area in graph theory,
see a recent monograph \cite{Li-Mao5} by Li and Mao on generalized connectivity of
undirected graphs, see also a survey paper \cite{Li-Mao} of the area.

To extend generalized $k$-connectivity to directed
graphs, note that an $S$-tree is a connected subgraph of $G$
containing $S$. In fact, in the definition of $\kappa_S(G)$ we could
replace ``an $S$-tree'' by ``a connected subgraph of $G$ containing
$S$.'' Therefore, we define {\em strong subgraph $k$-connectivity}
by replacing ``connected'' with ``strongly connected'' (or, simply, ``strong'') as follows.
Let $D=(V(D),A(D))$ be a digraph of order $n$, $S\subseteq V$ a
$k$-subset of $V(D)$ and $2\le k\leq n$.
Strong
subgraphs $D_1, \dots , D_p$ containing $S$ are said to be {\em $S$-internally disjoint} or, simply,
{\em internally disjoint} if $V(D_i)\cap V(D_j)=S$ and $A(D_i)\cap
A(D_j)=\emptyset$ for all $1\le i<j\le p$.

 Let $\kappa_S(D)$ be the maximum number of
internally disjoint strong digraphs containing $S$ in $D$. The {\em
strong subgraph $k$-connectivity} is defined as
$$\kappa_k(D)=\min\{\kappa_S(D)\mid S\subseteq V(D), |S|=k\}.$$
By definition, $\kappa_2(D)=0$ if $D$ is not strong.
Note that we define a digraph with one vertex to be strongly connected.
Strong subgraph $k$-connectivity allows us to extend applications of generalized $k$-connectivity  described in \cite{Li-Mao,Li-Mao5}
from undirected to directed graphs.


We will now overview results and conjectures on generalized $k$-connectivity related to
results and open problems of our paper.
Li, Li and Zhou \cite{Li-Li-Zhou}
showed that given a fixed positive integer $\ell$,
for any graph $G$ the problem of deciding whether $\kappa_3(G) \ge
\ell$ can be solved in polynomial time. This was generalized by Li
and Li \cite{Li-Li} who proved that given two fixed positive
integers $k\ge 2$ and $\ell$, for any graph $G$ the problem of deciding
whether $\kappa_k(G)\ge  \ell$ can be solved in polynomial time. For
a fixed integer  $k$, but an {\em arbitrary} (i.e. part of input) integer $\ell$, Li and Li
\cite{Li-Li} showed that the complexity changes provided P$\neq$NP:
Let $k\ge 4$ be a fixed integer. For a graph $G,$ a $k$-subset $S$ of $V (G)$ and an
integer $\ell$ ($\ell \ge 2$), it is NP-complete to decide whether
$\kappa_S(G)\ge \ell$. Solving a conjecture of S. Li \cite{LiThesis}, Chen, Li, Liu
and Mao \cite{Chen-Li-Liu-Mao} proved that in the above result, the bound 4 on $k$ can
be replaced by 3 (which is the best possible provided P$\neq$NP).
Note that another conjecture of S. Li \cite{LiThesis} remains
open  \cite{Li-Mao}: for a fixed integer $k\ge 3$, given a graph $G$ and an integer $\ell \ge 2$,
it is NP-complete to decide whether
$\kappa_k(G)\ge \ell .$ Thus, the ``global'' analog of the generalized local connectivity intractability result still remains
open. Li and Li \cite{Li-Li} proved an intractability result similar to that of \cite{Chen-Li-Liu-Mao} given above
when $\ell$ is
fixed but $k$ is {\em arbitrary}: For a graph $G$ and a subset $S$ of $V(G)$, it is NP-complete to decide whether $\kappa_S(G)\ge \ell$,
where $\ell\ge 2$ is a fixed integer.

It turns out that computing strong subgraph $k$-connectivity becomes intractable much
earlier with respect to $k$ and $\ell$ above.
Let $k \ge 2$ and $\ell\ge 2$ be fixed integers.
In Theorem~\ref{thm:2complexity} by reduction from the {\sc Directed 2-Linkage}
problem\footnote{The {\sc Directed $k$-Linkage} problem is formulated in the next section.} we prove that deciding whether
$\kappa_S(D)\ge \ell$ is NP-complete for a $k$-subset $S$ of $V(D)$.
Similarly to generalized $k$-connectivity,
we do not know whether the problem of deciding $\kappa_k(D)\ge \ell$ is NP-complete for fixed $k\ge 2$ and $\ell \ge 2$,
but we conjecture that it is the case.

Thomassen \cite{Thom} showed that for every positive integer $p$ there are digraphs which are strongly $p$-connected,
but which contain a pair of vertices not belonging to the same directed cycle. This implies that for every positive integer $p$ there are strongly $p$-connected
digraphs $D$ such that $\kappa_2(D)=1$.
Indeed, let $x$ and $y$ be vertices in a strongly $p$-connected  digraph $D$ such that no cycle contains both $x$ and $y$. Suppose $\kappa_2(D)\ge 2$. Then there are $\{x,y\}$-internally disjoint subgraphs $H_1$ and $H_2$ containing $x$ and $y$. But then a path from $x$ to $y$ in $H_1$ and a path from $y$ to $x$ in $H_2$ form a cycle in $D$, a contradiction.

The above negative results motivate
studying strong subgraph $k$-connectivity for special classes of digraphs.
Arguably the most studied of them is the class of
tournaments, see, e.g., a recent informative account
 \cite{Bang-Jensen-Havet} on tournaments and semicomplete digraphs by Bang-Jensen and Havet.
A digraph is {\em semicomplete} if there is at least one arc between
any pair of vertices. We show that the problem of deciding whether
$\kappa_k(D)\ge \ell$ for every semicomplete digraphs is
polynomial-time solvable for fixed $k$ and $\ell$ (Theorem~\ref{thm5}).
This result can be viewed as an analog of the corresponding result of Li and Li \cite{Li-Li} for
$\kappa_k(G)$. The main tool used in our proof is a recent {\sc Directed
$k$-Linkage} theorem\footnote{Another interesting recent result on  {\sc Directed
$k$-Linkage}  was published in \cite{Chud-Scott-SeymourAM}.} of Chudnovsky, Scott and Seymour
\cite{Chud-Scott-Seymour}.

A digraph $D$ is called {\em symmetric} if for every arc $xy$ there is an opposite arc $yx$.
Thus, a symmetric digraph $D$ can be obtained from its underlying undirected graph $G$ by replacing
each edge of $G$ with the corresponding arcs of both directions. We will say that $D$ is the {\em complete biorientation} of $G$ and
denote this by $D=\overleftrightarrow{G}.$
We will show that for any connected graph $G$, the parameter
$\kappa_2(\overleftrightarrow{G})$ can be computed in polynomial
time (Theorem~\ref{thm6}). This result is best possible in the following sense, unless P$=$NP.
Let $D$ be a symmetric digraph and $k\geq 3$ a fixed integer. Then it is NP-complete to decide
whether $\kappa_S(D)\geq \ell$ for $S\subseteq V(D)$ with $|S|=k$ (Theorem \ref{thm7}).
To prove Theorem \ref{thm7}, we use an NP-complete problem from
\cite{Chen-Li-Liu-Mao}. If we fix not only $k\ge 2$ but also $\ell\ge 2$, the complexity changes again (unless P$=$NP):
in Theorem \ref{thm:Sym}, we show that one can decide in polynomial time whether $\kappa_k(D)\geq \ell$.
To prove Theorem \ref{thm:Sym}, we use the celebrated result of
Robertson and Seymour \cite{RoSe} on the {\sc Undirected $p$-Linkage} problem.

Some inequalities concerning parameter $\kappa_k(G)$ were
obtained in the literature, see e.g. \cite{Li-Mao-Sun, Sun-Li}. For a
connected graph $G$ of order $n$, Li, Mao and Sun \cite{Li-Mao-Sun}
obtained the following inequality for $\kappa_k(G)$: $1\leq
\kappa_k(G)\leq n-\lceil \frac{k}{2}\rceil$, where $2 \leq k \leq n$. Moreover, the upper and lower
bounds are sharp. In the same paper, they also characterized graphs
$G$ with $\kappa_k(G)= n-\lceil \frac{k}{2}\rceil$.

Let $D$ be a strong digraph with order $n$. For $2\leq k\leq n$, we
prove that $1\leq \kappa_k(D)\leq n-1$  (Theorem~\ref{thm2}).
The
bounds are sharp; we also characterize those digraphs $D$ for which
$\kappa_k(D)$ attains the upper bound.
The main tool used in the proof of Theorem \ref{thm2} is a Hamiltonian cycle
decomposition theorem of Tillson \cite{Tillson}.

For a positive integer $m$, let $[m]=\{1,2,\dots ,m\}.$

The paper is organized as follows. The next section is devoted NP-completeness results and polynomial algorithms discussed above. In Section \ref{sec:bounds} we prove sharp lower and upper bounds on strong subgraph $k$-connectivity also discussed above. We conclude the paper with Section \ref{sec:openq}, where we discuss further direction of research on strong subgraph $k$-connectivity and state some open problems.


\section{Algorithms and Complexity}

It is easy to decide whether $\kappa_S(D)\ge 1$ for a digraph $D$: it holds if and only if $D$ is strong.
Unfortunately, deciding whether $\kappa_S(D)\ge 2$ is already NP-complete for $S \subseteq V(D)$
with $|S|=k$, where $k\ge 2$ is a fixed integer.

The well-known {\sc Directed $q$-Linkage} problem \cite{Bang-Jensen-Gutin} is of interest in the next three theorems.
The problem is formulated as follows:
for a fixed integer $q\ge 2$, given a digraph $D$ and a
(terminal) sequence $((s_1,t_1),\dots ,(s_q,t_q))$ of  distinct vertices
of $D,$ decide whether $D$ has $q$ vertex-disjoint paths $P_1,\dots
,P_q$, where $P_i$ starts at $s_i$ and ends at $t_i$ for all $i\in [q].$

Let us prove our main intractability result.

\begin{thm}\label{thm:2complexity} Let $k\ge 2$ and $\ell\ge 2$ be fixed integers.
Let $D$ be a digraph and $S \subseteq V(D)$ with $|S|=k$. The
problem of deciding whether $\kappa_S(D)\ge \ell$ is NP-complete.
\end{thm}
\begin{pf}
Clearly, the problem is in NP. To show it is NP-hard, we reduce from
the {\sc Directed 2-Linkage} problem, which is NP-complete
\cite{Fortune-Hopcroft-Wyl}.

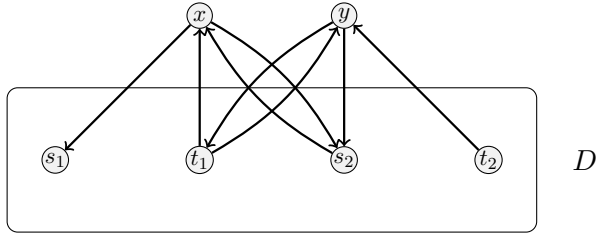
\begin{figure}[tb]
\begin{center}
\tikzstyle{vertexX}=[circle,draw, fill=gray!10, minimum size=12pt, scale=0.8, inner sep=0.3pt]
\begin{tikzpicture}[scale=0.64]
\node (x) at (4.0,4.0) [vertexX] {$x$};
\node (y) at (7.0,4.0) [vertexX] {$y$};
\node (s1) at (1.0,1.0) [vertexX] {$s_1$};
\node (t1) at (4.0,1.0) [vertexX] {$t_1$};
\node (s2) at (7.0,1.0) [vertexX] {$s_2$};
\node (t2) at (10.0,1.0) [vertexX] {$t_2$};
\draw [->, line width=0.03cm] (x) -- (s1);
\draw [->, line width=0.03cm] (t1) -- (x);
\draw [->, line width=0.03cm] (y) -- (s2);
\draw [->, line width=0.03cm] (t2) -- (y);

\draw [->, line width=0.03cm] (x) to [out=330, in=120] (s2);
\draw [->, line width=0.03cm] (s2) to [out=150, in=300] (x);

\draw [->, line width=0.03cm] (y) to [out=210, in=60] (t1);
\draw [->, line width=0.03cm] (t1) to [out=30, in=240] (y);

\draw [rounded corners] (0,-0.5) rectangle (11,2.5);
\node at (12.0,1.0) {$D$};
 \end{tikzpicture}
\end{center}
\caption{The digraph $D'$.} \label{picDp}
\end{figure}

 Let us first consider the case of $\ell=2$ and $k=2$.
Let $(D,s_1,t_1,s_2,t_2)$ be an instance of  {\sc Directed
2-Linkage}. Let us construct a new digraph $D'$ (see Figure~\ref{picDp}) by adding to $D$ vertices $x,y$ and arcs
$$t_1x,xs_1, t_2y,ys_2,  xs_2,s_2x,yt_1,t_1y.$$
Let $S=\{x,y\}.$
It remains to show that $(D,s_1,t_1,s_2,t_2)$ is a positive
instance of {\sc Directed 2-Linkage} if and only if $\kappa_S(D')\ge
2$.

Let $(D,s_1,t_1,s_2,t_2)$ be a positive instance of  {\sc Directed
2-Linkage} with vertex-disjoint paths $P_1,P_2$ from $s_1$ to $t_1$
and from $s_2$ to $t_2$, respectively. Then there are two internally
disjoint strong subgraphs containing $S$ of $D'$, one induced
by the arcs of $P_1$ and $t_1x,xs_1,t_1y,yt_1$ and the other by the arcs of $P_2$
and $t_2y, ys_2, xs_2,s_2x$.

Let $D'$ have two internally disjoint strong subgraphs $H_1,H_2$ containing $S$.
Since the in-degree of $x$ in $D'$ is $2$, we may without loss of generality assume that
$t_1 \in V(H_1)$ and $s_2 \in V(H_2)$. As $y$ has in-degree $2$ and $t_1 \in V(H_1)$ we must
have $t_2 \in V(H_2)$. As the out-degree of $x$ is $2$, we analogously have $s_1 \in V(H_1)$
(as $s_2 \in V(H_2)$). So, for $i=1,2$, both $s_i$ and $t_i$ are in $H_i$.
Therefore, there must be a path $P_i$ from $s_i$ to $t_i$ in $H_i$
and by definition of $D'$, $P_i$ will not have vertices outside of $D$.
As $H_1$ and $H_2$ are internally disjoint, the paths are disjoint.

Now let us consider the case of $\ell \ge 3$ and $k=2$.
Add to $D'$ $\ell -2$ copies of the 2-cycle $xyx$ and subdivide the arcs of every copy to avoid parallel arcs.
Let us denote the new digraph by $D''$.
Assume that there are $\ell$ internally disjoint strong subgraphs, $H_1,H_2,\ldots H_{\ell}$, containing $S$ in $D''$.
As the out-degree of $y$ in $D''$ is $\ell$ we can without loss of generality assume that $t_1 \in V(H_1)$,
$s_2 \in V(H_2)$ and the $\ell -2$ (subdivided) arcs from $y$ to $x$ belong to $H_3,H_4,\ldots,H_{\ell}$, respectively.
As $t_1 \in V(H_1)$ and the in-degree of $y$ is $\ell$ no (subdivided) arc from $x$ to $y$ belongs to $H_1$.
Analogously, since $s_2 \in V(H_2)$ and the out-degree of $x$ is $\ell$, no (subdivided) arc from $x$ to $y$ belongs to $H_2$.
Therefore the (subdivided) arcs from $x$ to $y$ belong to $H_3,H_4,\ldots,H_{\ell}$, respectively.
As in the case when $\ell=2$ we now note that $s_1 \in V(H_1)$ and $t_2 \in V(H_2)$ and that there therefore exists
disjoint paths from $s_1$ to $t_1$ and $s_2$ to $t_2$ in $D$, respectively.

Conversely if there exists disjoint paths from $s_1$ to $t_1$ and $s_2$ to $t_2$ in $D$, then it is not difficult to
create $\ell$ internally disjoint strong subgraphs containing $S$ in $D''$ using the same approach as when
$\ell=2$ as each (subdivided) $2$-cycle $xyx$ also gives rise to a strong subgraph containing $S$.
Thus, we have proved the theorem in the case of $k=2$ and $\ell \ge 2$.

It remains to consider the case of $\ell \ge 2$ and $k\ge 3$. Add to $D''$ (where $D''=D'$ for $\ell =2$) $k-2$ new vertices $x_1,\dots ,x_{k-2}$ and arcs of $\ell$ 2-cycles $xx_ix$ for each $i\in [k-2]$.
Subdivide the new arcs to avoid parallel arcs. Let $S=\{x,y,x_1,\dots ,x_{k-2}\}$. It is not hard to see that the resulting digraph has $\ell$ internally disjoint strong subgraphs if and only if $(D,s_1,t_1,s_2,t_2)$ is a positive instance of  {\sc Directed 2-Linkage}.
\end{pf}

Recently, Chudnovsky, Scott and Seymour
\cite{Chud-Scott-Seymour} proved the following powerful result,
which  was already used in \cite{Bang-Jensen-Larsen-Maddaloni}.

\begin{thm}\label{thm:CSS}\cite{Chud-Scott-Seymour}
Let $q$ and $c$ be fixed positive integers. Then the  {\sc Directed
$q$-Linkage} problem on a digraph $D$ whose vertex set can be
partitioned into $c$ sets each inducing a semicomplete digraph and
a terminal sequence $((s_1,t_1),\dots ,(s_q,t_q))$ of  distinct vertices
of $D$, can be solved in polynomial time.
\end{thm}

Now we will consider the problem of deciding whether $\kappa_k(D)\ge \ell$
for a semicomplete digraph $D$. We will first prove the following:

\begin{lem}\label{lemX1}
Let $k$ and $\ell$ be fixed positive integers. Let $D$ be a digraph
and let $X_1,X_2,\ldots,X_{\ell}$ be $\ell$ vertex
disjoint subsets of $V(D)$, such that $|X_i| \leq k$ for all $i\in
[\ell]$. Let $X = \cup_{i=1}^{\ell} X_i$ and assume that every
vertex in $V(D) \setminus X$ is adjacent to every other vertex in
$D$. Then we can in polynomial time decide if there exists vertex
disjoint subsets $Z_1,Z_2,\ldots,Z_{\ell}$ of $V(D)$, such that $X_i
\subseteq Z_i$ and $D[Z_i]$ is strongly connected for each $i\in
[\ell]$.
\end{lem}

\begin{pf}
 Let $C_i^1, C_i^2, \ldots, C_i^{r_i}$ be the strongly connected components in $D[X_i]$, such that
there is no arc from $C_i^b$ to $C_i^a$ for $1 \leq a < b \leq r_i$.
We consider the following two cases.

\2

{\bf Case 1:} {\em $D[X_i]$ has a unique initial and a unique terminal component (which can be the same component)
for all $i=1,2,\ldots,\ell$.}

Let ${\cal T} = \emptyset$. For each $i=1,2,\ldots,\ell$, do the
following.  If $D[X_i]$ is strongly connected then set $Z_i=X_i$ and
delete $X_i$ from $D$. Otherwise, contract every strong component
$C_i^j$ to a vertex $c_i^j$ and look at all possible subsequences,
of $c_i^1,c_i^2,\ldots,c_i^{r_i}$ which start with $c_i^{r_i}$ and
end with $c_i^1$. Let $Z=(z_1,z_2,\ldots,z_r)$ be such a
subsequence, where $z_1=c_i^{r_1}$, $z_r=c_i^1$ and $2 \leq r \leq
r_i$. Now duplicate every vertex $z_a$ to $z_a^s$ and $z_a^t$, for
all $a=2,3,\ldots,r-1$ and remove every $c_i^j$ that does not appear
in the subsequence. We now add the sequence ${\cal T}_i=((c_i^{r_1}
z_2^t), (z_2^s,z_3^t), (z_3^s,
z_4^t), \ldots, (z_{r-1}^s, c_i^1))$ to our terminal sequence ${\cal
T}$.

After having done the above for all $i=1,2,\ldots,\ell$ we use
Theorem~\ref{thm:CSS} in order to determine if there are vertex disjoint paths satisfying our
terminal sequence ${\cal T}$ (that is, for every $(s,t) \in {\cal T}$ there is a path from $s$ to $t$).
If such a linkage exists (for the terminal sequence ${\cal T}$ of some subsequences above)
then let $Z_i$ include all internal vertices on paths between the pairs of vertices in ${\cal T}_i$ as well as $X_i$ itself.
Now observe that $D[Z_i]$ is strongly connected and all $Z_1,Z_2,\ldots,Z_{\ell}$ are vertex disjoint,
as desired.

We will now show that if there exists $Z_i$, such that $D[Z_i]$ is strongly connected and all $Z_i$ are vertex
disjoint, then there exists a desired linkage.  So, assume that such $Z_i$ exist.  As $D[Z_i]$ is strong, we note that
it remains strong after contracting all strong components of $D[X_i]$ to vertices.  Thereafter there exists a
shortest path $P$ from the terminal strong component of $D[X_i]$ to the initial strong component of $D[X_i]$.
Let the vertices on $P$ which correspond to (contracted) strong components of $D[X_i]$ be $(z_1,z_2,\ldots,z_r)$
(in the order they appear on $P$) and  using this as the subsequence in our algorithm the subpaths of $P$ gives us
the desired linkage between the $z_i$'s.
Doing the above for all $i=1,2,\ldots,\ell$ we see that our algorithm will indeed find the desired linkage (when considering
the subsequences constructed above).

As $k$ and $\ell$ are constants, we note that there are at most a constant number of subsequences to consider, so the
algorithm runs in polynomial time. This completes Case 1.

\2

{\bf Case 2:} {\em Case 1 does not hold.}

We will in this case transform the problem, such that we can solve it using Case~1.
For all $i=1,2,\ldots,\ell$ proceed as follows. Initiate a set $Q$ as an empty set.
If there is a unique initial strong component in $D[X_i]$ and a unique strong terminal component in $D[X_i]$ then let $X_i'=X_i$.
If this is not the case, then let $I=\{I_1,I_2,\ldots,I_p\}$ denote the set of initial strong components in $D[X_i]$ and
let $T=\{T_1,T_2,\ldots,T_q\}$ denote the set of terminal strong components in $D[X_i]$.
For every $I_a \in I$ choose a vertex, $v_a \in V(D) \setminus (X_1 \cup X_2 \cup \cdots \cup X_{\ell}\cup Q)$ such that $v_a$
has at least one arc into the component $I_a$. We allow repetition of vertices in the sequence $v_1,v_2,\ldots,v_p$.
(Such vertices $v_j$ must exist if there is a set $Z_i$ containing $X_i$ such that $D[Z_i]$ is strong.) 
Analogously, for each  $T_b \in T$ choose a vertex, $w_b \in V(D) \setminus (X_1 \cup X_2 \cup \cdots \cup X_{\ell}\cup Q)$ such that $w_b$
has at least one arc into it from the component $T_b$. Again we allow $w_1,w_2,\ldots,w_q$ to be not necessarily distinct.
Now add vertices of $v_1,v_2,\ldots,v_p$ and $w_1,w_2,\ldots,w_q$ to $Q$.

If for some $i$ we cannot choose $v_1,v_2,\ldots,v_p$ as above, we stop and consider other choices
for the previous values of $i$. Analogously, for $w_1,w_2,\ldots,w_q$.

If we have succeeded in choosing
$v_1,v_2,\ldots,v_p$ and $w_1,w_2,\ldots,w_q$ for every $i\in
[\ell]$, then for each $i\in [\ell]$ we add the corresponding
vertices $v_1,v_2,\ldots,v_p$ and $w_1,w_2,\ldots,w_q$ to $X_i$ and
call the resulting set $X_i'$. Note that $|X_i'| \leq |X_i|+(p+q)
\leq 2k$.

If $C$ is a terminal component in $D[X_i']$ then $C$ must contain a vertex not in $X_i$, as otherwise
$C$ would be a terminal component of $D[X_i]$ a contradiction to $X_i'$ containing a vertex (not in $X_i$ and therefore not in $C$)
that has an arc into it from $C$.
However, as all vertices not in $X_i$ are adjacent, this implies that there is a unique terminal strong component in $D[X_i']$.
Analogously, there is a unique initial strong component in $D[X_i']$.

We now use the approach in Case~1, for all possible choices of vertices $v_j$ and $w_j$ for all $i\in [\ell]$.
As there are at most $n^k$ possible choices of vertices $v_j$ and $w_j$ for each $i$ observe that we have to use
the approach in Case~1 at most $n^{k\ell}$ times, which is a polynomial as $k$ and $\ell$ are constants.

If the above algorithm finds the $\ell$ sets, $Z_1,Z_2,\ldots,Z_{\ell}$, then clearly they exist.
Conversely, if the sets do exist then when $D[X_i]$ is not strong, observe that each initial strong component in $D[X_i]$ must
have an arc into it from a vertex in $Z_i$ and each terminal strong component in $D[X_i]$ must
have an arc into out of it to a vertex in $Z_i$. Picking these vertices as our vertices $v_j$ and $w_j$, observe that
our algorithm will indeed find sets $Z_1,Z_2,\ldots,Z_{\ell}$, as desired.
\end{pf}

\begin{thm}\label{thm5}
For any fixed integers $k, \ell \ge 2$, we can decide whether
$\kappa_k(D)\ge \ell$ for a semicomplete digraph $D$ in polynomial
time.
\end{thm}
\begin{pf}
Let $k, \ell \ge 2$ be fixed and let $S=\{s_1,\dots ,s_k\}$ be a set of
vertices of a semicomplete digraph $D$. To prove this theorem it
suffices to show that deciding whether $\kappa_S(D)\ge \ell$ can be
done in polynomial time.

Let $D^*$ be obtained from $D$ by replacing every $s_i$ by $\ell$ copies,
i.e. replacing $s_i$ with $S_i=\{x_i^1,x_i^2,\ldots,x_i^{\ell}\}$ for all $i=1,2,\ldots,k$.
Let $X_i=\{x_1^i,x_2^i,\ldots,x_k^i\}$ for all $i=1,2,\ldots,\ell$.
Assume for now that $X_i$ is an independent set in $D^*$ (this will change later),
but if $s_i y $ is an arc from $S$ to $V(D) \setminus S$, then $x_i^a y$ is in $D^*$ for
all $a=1,2,\ldots,\ell$.
An analogously, if $y s_i$ is an arc from $V(D) \setminus S$ to $S$, then $y x_i^a$ is in $D^*$ for
all $a=1,2,\ldots,\ell$.

Let $A_1,A_2,\ldots,A_{\ell}$ be a partition of the arcs in $D[S]$,
where some sets may be empty. That is, every arc in $D[S]$ belongs
to exactly one $A_i$. For each $i=1,2,\ldots,\ell$ add the arcs of
$A_i$ to $D^*[X_i]$. That is, if $s_as_b \in A_i$ then add the arc
$x_a^i x_b^i$ to $D^*$. This
completes the construction of $D^*$ (for a given partition
$A_1,A_2,\ldots,A_{\ell}$).

We can now decide if there exist disjoint vertex sets $Z_i$ in $D^*$ such that
$X_i \subseteq Z_i$ and $D^*[Z_i]$ is strongly connected for all $i=1,2,\ldots,\ell$ in polynomial
time by Lemma~\ref{lemX1}. If, for some partition, $A_1,A_2,\ldots,A_{\ell}$, such $Z_i$'s exist
then we will show that $\kappa_S(D) \ge \ell$ and if this is not the case then we will show that $\kappa_S(D) < \ell$.
As there are only a polynomial number of partitions $A_1,A_2,\ldots,A_{\ell}$ (as $\ell$ and $k$ are constants),
this gives us a polynomial algorithm.

First assume that such $Z_i$'s exist for some partition, $A_1,A_2,\ldots,A_{\ell}$.
Then the subgraph in $D$ on vertex set $Z_i$ and with the arcs $(A(D[Z_i])\setminus A(D[S])) \cup A_{\ell}$ is strongly
connected and as all $Z_i$'s are vertex disjoint (and the arc sets $A_i$'s are disjoint) observe that
$\kappa_S(D) \ge \ell$, as desired.

Conversely if $\kappa_S(D) \ge \ell$, then there exists strongly connected subgraphs $Y_1,Y_2,\ldots,Y_{\ell}$
such that $V(Y_i) \cap V(Y_j) = S$ for all $i \not= j$. Without loss of generality, we may assume that every arc of $D[S]$
belongs to some $Y_i$ (as otherwise just add it to some $Y_i$). Letting $A_i=Y_i[S]$ and $Z_i=V(Y_i)$ observe that our
algorithm does find the desired $Z_i$'s and we are done.
\end{pf}

Now we turn our attention to symmetric graphs. We start with the following structural result.

\begin{thm}\label{thm6}
For every graph $G$ we have $\kappa_2(\overleftrightarrow{G})=\kappa(G)$.
\end{thm}
\begin{pf}
We may assume that $G$ is a connected graph.
Let $D$ be a digraph whose  underlying undirected graph is $G$ and let $S=\{x,y\}$, where $x,y$ are distinct vertices of $D$.
Observe that $\kappa_S(G)\ge \kappa_S(D)$. Indeed, let $p=\kappa_S(D)$ and let $D_1,\dots ,D_p$ be $S$-internally disjoint  strong subgraphs of $D$.
Thus,  by choosing a path from $x$ to $y$ in each $D_i$, we obtain $p$ internally disjoint paths from $x$ to $y$, which correspond to $p$ internally disjoint paths between $x$ and $y$ in $G$. Thus, $\kappa (G)\ge \kappa_2(D)$ and
it suffices to show that $\kappa_S(\overleftrightarrow{G})\geq
\kappa(G)$.

Let
$\kappa_S(\overleftrightarrow{G})=\kappa_2(\overleftrightarrow{G})$
for some $S=\{x, y\}\subseteq V(\overleftrightarrow{G})$. We know that
there are at least $\kappa(G)$ internally disjoint paths connecting
$x$ and $y$ in $G$, say $P_i~(i\in [\kappa(G)])$. For each
$i\in [\kappa(G)]$, we can obtain a strong subgraph containing
$S$, say $D_i$, in $\overleftrightarrow{G}$ by replacing each edge
of $P_i$ with the corresponding arcs of both directions. Clearly,
any two such subgraphs are internally disjoint, so we have
$\kappa_2(\overleftrightarrow{G})=\kappa_S(\overleftrightarrow{G})\geq
\kappa(G)$ and we are done.
\end{pf}

Theorem~\ref{thm6} immediatly implies the following positive result, which follows from
the fact that $\kappa(G)$ can be computed in polynomial time.

\begin{cor}\label{cor6}
For a graph $G$, $\kappa_2(\overleftrightarrow{G})$ can be computed in polynomial time.
\end{cor}

Theorem~\ref{thm6} states that $\kappa_k(\overleftrightarrow{G})=\kappa_k(G)$ when $k=2$.
However when $k \geq 3$, then $\kappa_k(\overleftrightarrow{G})$ is not always equal to $\kappa_k(G)$,
as can be seen by $\kappa_3(\overleftrightarrow{K_3}) = 2 \not= 1 = \kappa_3(K_3)$.
Chen, Li, Liu and Mao \cite{Chen-Li-Liu-Mao} introduced the
following problem, which turned out to be NP-complete.

\2

{\sc CLLM Problem:} Given a tripartite graph $G=(V, E)$ with a
3-partition $(\overline{U}, \overline{V}, \overline{W})$ such that
$|\overline{U}|=|\overline{V}|=|\overline{W}|=q$, decide whether
there is a partition of $V$ into $q$ disjoint 3-sets $V_1,
\dots, V_q$ such that for every $V_i= \{v_{i_1}, v_{i_2}, v_{i_3}\}$
$v_{i_1} \in \overline{U}, v_{i_2} \in \overline{V},
v_{i_3} \in \overline{W}$ and $G[V_i]$ is connected.

\begin{lem}\label{thm02}\cite{Chen-Li-Liu-Mao} The CLLM Problem is NP-complete.
\end{lem}

Now restricted to symmetric digraphs $D$, for any fixed integer $k\geq
3$, the problem of deciding whether $\kappa_S(D)\geq \ell~(\ell \geq
1)$ is NP-complete for $S\subseteq V(D)$ with $|S|=k$.

\begin{thm}\label{thm7}
For any fixed integer $k\geq 3$, given a symmetric digraph $D$, a
$k$-subset $S$ of $V(D)$ and an integer $\ell~(\ell \geq 1)$,
deciding whether $\kappa_S(D)\geq
\ell$, is NP-complete.
\end{thm}
\begin{pf}
It is easy to see that this problem is in NP. We divide our proof
into two steps:

In the first step, let $G$ be a tripartite graph with 3-partition
$(\overline{U}, \overline{V}, \overline{W})$ such that
$|\overline{U}|=|\overline{V}|=|\overline{W}|=q$. We will construct
a graph $H$, a $k$-subset $S\subseteq V(H)$ and an integer
$\ell$ such that there are $\ell$ internally disjoint $S$-trees in
$H$ if and only if $G$ is a positive instance of the CLLM Problem.

We define $H$ as follows: let $V(H)=V(G)\cup \{x_j\mid 1\leq j\leq
k\}$ and $E(H)=E(G)\cup \{x_ju\mid 1\leq j\leq k-2, u\in
\overline{U}\}\cup \{x_{k-1}v\mid v\in
\overline{V}\} \cup \{x_{k}w\mid w\in
\overline{W}\}$. Set $S=\{x_j\mid 1\leq j\leq k\}$ and $\ell=q$.

If there are $\ell$ internally disjoint $S$-trees in $H$, then each
tree contains exactly a vertex from $\overline{U}$, a vertex from
$\overline{V}$ and a vertex from $\overline{W}$ since
${\rm deg}_H(x_i)=\ell$ for all $i\in [k]$. Furthermore, in each such
tree, elements of $\{u_i\mid 1\leq i\leq k-2\}$ have exactly one
common neighbor in $\overline{U}$. Since these $\ell$ trees are
internally disjoint, there is a partition of $V(G)$ into
$q=\ell$ disjoint sets $V_1, V_2, \cdots, V_q$ each having three
vertices, such that for every $V_i= \{v_{i_1}, v_{i_2}, v_{i_3}\}$
we have that $v_{i_1} \in \overline{U}, v_{i_2} \in \overline{V},
v_{i_3} \in \overline{W}$, and $G[V_i]$ is connected.

If there is a partition of $V(G)$ into $q=\ell$ disjoint sets
$V_1, V_2, \cdots, V_q$ each having three vertices, such that for every
$V_i= \{v_{i_1}, v_{i_2}, v_{i_3}\}$ we have $v_{i_1} \in
\overline{U}, v_{i_2} \in \overline{V}, v_{i_3} \in \overline{W}$,
and $G[V_i]$ is connected, then let $T_i$ be a spanning tree of
$G[V_i]$ together with the edge set $\{x_jv_{i_1}\mid 1\leq j\leq
k-2\}\cup\{u_{k-1}v_{i_2}\}\cup\{u_{k}v_{i_3}\}$, where
$V_i=\{v_{i_1}, v_{i_2}, v_{i_3}\}$. It is easy to see that $T_1,
T_2, \cdots, T_{\ell}$ are the desired internally disjoint $S$-trees.

In the second step, we construct a symmetric digraph $D$ from $H$ by
replacing each edge with the corresponding arcs of both directions.
If there are $\ell$
internally disjoint $S$-trees in $H$, then for each such tree, we
can get a strong subgraph containing $S$ in $D$ by replacing each edge
with the corresponding arcs of both directions. Clearly, all these
subgraphs of $D$ are internally disjoint and contain $S$.
 If there are $\ell$ internally disjoint strong subgraphs
containing $S$, say $D_i~(1\leq i\leq \ell)$, in $D$, then each
$D_i$ contains exactly a vertex from $\overline{U}$, a vertex from
$\overline{V}$ and a vertex from $\overline{W}$ since
$|\overline{U}|=|\overline{V}|=|\overline{W}|=\ell$. For every $
i\in [\ell]$, let $T_i$ be a spanning tree of the underlying undirected
graph of $D_i$. Observe that $T_1,\dots ,T_{\ell}$ are internally disjoint
$S$-trees in $H$.
We now have that there are $\ell$ internally disjoint $S$-trees in
$H$ if and only if there are $\ell$ internally disjoint strong
subgraphs containing $S$ in $D$.

Now, by Lemma \ref{thm02} and the two steps above, we are done.
\end{pf}

The last theorem assumes that $k$ is fixed but $\ell$ is a part of input. When both $k$ and $\ell$ are fixed, the problem of deciding whether
$\kappa_S(D) \geq \ell$ for a symmetric digraph $D$, is polynomial-time solvable. We will start with the following technical lemma.

\begin{figure}[tb]
\begin{center}
\tikzstyle{vertexX}=[circle,draw, fill=gray!10, minimum size=13pt, scale=0.85, inner sep=0.6pt]
\begin{tikzpicture}[scale=0.24]
\node (s1) at (1.0,2.0) [vertexX] {$s_1$};
\node (s2) at (4.0,2.0) [vertexX] {$s_2$};
\node (s3) at (7.0,2.0) [vertexX] {$s_3$};
\node (s4) at (10.0,2.0) [vertexX] {$s_4$};
\node (s5) at (13.0,2.0) [vertexX] {$s_5$};
\node (s6) at (16.0,2.0) [vertexX] {$s_6$};
\node (s7) at (19.0,2.0) [vertexX] {$s_7$};
\node (x1) at (13.0,12.0) [vertexX] {$x_1$};
\node (x2) at (16.0,12.0) [vertexX] {$x_2$};
\node (x3) at (1.0,7.0) [vertexX] {$x_3$};
\node (x4) at (4.0,7.0) [vertexX] {$x_4$};
\node (x5) at (8.5,7.0) [vertexX] {$x_5$};
\node (x6) at (13.0,7.0) [vertexX] {$x_6$};
\node (x7) at (16.0,7.0) [vertexX] {$x_7$};
\node (x8) at (19.0,7.0) [vertexX] {$x_8$};
\draw [line width=0.05cm] (s1) -- (x3);
\draw [line width=0.05cm] (s2) -- (x4);
\draw [line width=0.05cm] (s3) -- (x5);
\draw [line width=0.05cm] (s4) -- (x5);
\draw [line width=0.05cm] (s4) to [out=80, in=240] (x1);
\draw [line width=0.05cm] (s5) -- (x6);
\draw [line width=0.05cm] (s6) -- (x7);
\draw [line width=0.05cm] (s7) -- (x8);
\draw [line width=0.05cm] (x3) -- (x4);
\draw [line width=0.05cm] (x1) -- (x2);
\draw [line width=0.05cm] (x1) -- (x6);
\draw [line width=0.05cm] (x2) -- (x8);
\draw [line width=0.05cm] (x7) -- (x8);
\draw [rounded corners] (-0.5,0.5) rectangle (20.5,3.5);
\node at (10.0,-1) {The forest $F_i$};
 \end{tikzpicture} \hfill
\begin{tikzpicture}[scale=0.24]
\node (s1) at (1.0,2.0) [vertexX] {$s_1$};
\node (s2) at (4.0,2.0) [vertexX] {$s_2$};
\node (s3) at (7.0,2.0) [vertexX] {$s_3$};
\node (s4) at (10.0,2.0) [vertexX] {$s_4$};
\node (s5) at (13.0,2.0) [vertexX] {$s_5$};
\node (s6) at (16.0,2.0) [vertexX] {$s_6$};
\node (s7) at (19.0,2.0) [vertexX] {$s_7$};
\node (x1) at (13.0,12.0) [vertexX] {$x_1$};
\node (x2) at (16.0,12.0) [vertexX] {$x_2$};
\node (x3) at (1.0,7.0) [vertexX] {$x_3$};
\node (x4) at (4.0,7.0) [vertexX] {$x_4$};
\node (x5) at (8.5,7.0) [vertexX] {$x_5$};
\node (x6) at (13.0,7.0) [vertexX] {$x_6$};
\node (x7) at (16.0,7.0) [vertexX] {$x_7$};
\node (x8) at (19.0,7.0) [vertexX] {$x_8$};

\draw [line width=0.05cm] (s1) -- (s2);
\draw [line width=0.05cm] (s3) -- (s4);
\draw [line width=0.05cm] (s4) to [out=80, in=240] (x1);
\draw [line width=0.05cm] (s5) to [out=65, in=295] (x1);
\draw [line width=0.05cm] (s6) -- (x8);
\draw [line width=0.05cm] (s7) -- (x8);
\draw [line width=0.05cm] (x1) -- (x8);
\draw [rounded corners] (-0.5,0.5) rectangle (20.5,3.5);
\node at (10.0,-1) {The skeleton of $F_i$};
 \end{tikzpicture}
\end{center}
\caption{An example of the skeleton of a forest $F_i$, where $S=\{s_1,s_2,s_3,s_4,s_5,s_6,s_7\}$.} \label{figT}
\end{figure}
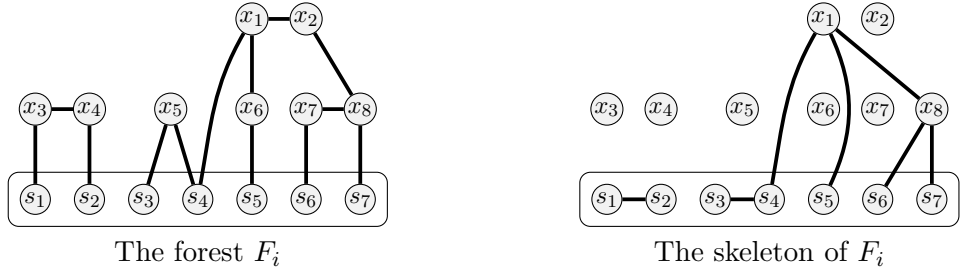

\begin{lem}\label{lemY}
Let $k,\ell \geq 2$ be fixed. Let $G$ be a graph and let $S \subseteq V(G)$ be an independent set in $G$ with $|S|=k$.
For $i\in [\ell]$, let $D_i$ be any set of arcs with both end-vertices in $S$. Let a forest $F_i$ in $G$
be called $(S,D_i)$-{\em acceptable} if the digraph $\overleftrightarrow{F_i}+D_i$ is strong and contains $S$.
In polynomial time, we can decide whether there exists edge-disjoint forests $F_1,F_2,\ldots,F_{\ell}$ such that $F_i$ is
$(S,D_i)$-acceptable for all $i\in [\ell]$ and $V(F_i) \cap V(F_j) \subseteq S$ for all $1 \leq i < j \leq \ell$.
\end{lem}
\begin{pf}
Assume that there exists a set ${\cal F}=\{F_1,F_2,\ldots,F_{\ell}\}$ of required forests. Observe that if there is a leaf $v\not\in S$ in a forest $F_i$, $v$ can be deleted from $F_i$
and $\cal F$ will remain the required set. Thus, we may assume that all leaves in $\cal F$ are vertices of $S$.

Below we will use the fact that in a tree without degree-2 vertices,
the number of internal vertices is smaller than the number of
leaves. This fact can be easily proved by induction by deleting a
leaf. Let $T$ be a tree in a forest of 
${\cal F}$ and let $T'$ be the tree obtained from
$T$ by contracting all degree-2 vertices not belonging to $S$. We
will call $T'$ the {\em skeleton} of $T$. Note that $T'$ may contain
edges, that are not edges of $G$ (see Figure~\ref{figT} for an
example). By producing the skeleton of every tree of $F_i$, we
obtain the {\em skeleton} of the forest $F_i$.

We will now bound the number of possible skeletons obtained from $F_i$'s.  By Cayley's formula, the number of distinct trees
on $n$ labeled vertices is bounded by $n^{n-2}$. By considering every tree on $n_T = 2|S|-1=2k-1$ vertices, then assigning the $n_T$
vertices to vertices of $G$ and finally deleting a subset of edges in the tree, we obtain a forest in $G$.  Note that
after deleting isolated vertices every skeleton obtained from an $F_i$ is created this way.  Therefore the number
of  possible skeletons is bounded by

\[
  n_F = (n_T)^{n_T-2} \times |V(G)|^{n_T} \times 2^{n_T-1}.
\]

Note that the above number is a polynomial in $|V(G)|$ as $k$ is considered constant, which implies that $n_T$ is constant.
Thus, the number of distinct skeletons for the set $\cal F$ is bounded by $n_F^{\ell}$, which is still a
polynomial in $|V(G)|$ as $\ell$ is also considered to be a constant.

We now consider a set of skeletons ${\cal F}'=\{F_1',F_2',\ldots,F_{\ell}'\}$ where
$F_i'$ is $(S,D_i)$-acceptable and $V(F_i') \cap V(F_j') \subseteq S$ for all $1 \leq i < j \leq \ell$.
It remains to show that there is a polynomial-time algorithm for deciding whether there exists a set ${\cal Q}=\{Q_1,Q_2,\ldots, Q_{\ell}\}$
of required forests such that $F'_i$ is the skeleton of $Q_i$. To obtain such an algorithm, we will use
the celebrated result of Robertson and Seymour \cite{RoSe}  that the {\sc Undirected $p$-Linkage} problem is
polynomial-time solvable.
For every forest $F_i' \in {\cal F}'$ and every $x \in V(F_i')$ make $d_{F_i'}(x)$ copies of $x$ in $G$, such that all copies have
the same neighbourhood as $x$. If $x$ belongs to several forests in ${\cal F}'$ (which can happen if $x \in S$)
then do the above for every forest, implying that we increase the number of copies of $x$ several times.
Let $U_x$ denote all copies of $x$.
Note that if $xy \in E(G)$ then all vertices in $U_x$ are adjacent to all vertices in $U_y$.
For every edge $uv$ in a forest in ${\cal F}'$ we now wish to find a path from a copy of $u$ to a copy of $v$. We can
decide if all such vertex-disjoint paths exist by the {\sc Undirected $p$-Linkage} problem.

One can show that at most one edge from $U_x$ to $U_y$ is used in such a collection of paths,
however this property can also be guaranteed by for each edge $uv$ adding a new vertex $z_{uv}$ and adding
all edges from $U_x \cup U_y$ to $z_{xy}$ instead of the edges between $U_x$ and $U_y$ (then at most one path of length at most $2$ from $U_x$ to
$U_y$ can be used, as it has to go through $z_{xy}$).

Therefore, if all of these vertex-disjoint paths exist, then they give us the desired set ${\cal Q}$ and if they do not exist
then the desired set ${\cal Q}$ does not exist.
\end{pf}



Now we can prove the following:

\begin{thm}\label{thm:Sym}
Let $k,l \geq 2$ be fixed. For any symmetric digraph $D$ and $S \subseteq V(D)$ with $|S|=k$ we can in polynomial
time decide whether $\kappa_S(D) \geq \ell$.
\end{thm}
\begin{pf}
Let $k$, $\ell$, $D$ and $S$ be defined as in the statement of the theorem.
Let $A_S$ be the set of arcs in $D[S]$. As $|S|=k$ we note that $|A_S| \leq 2{k \choose 2}$.

Let ${\cal P}=\{P_1,P_2,\ldots,P_{\ell}\}$ be any partition of $A_S$ (i.e., all sets of ${\cal P}$ are disjoint and their union is $A_S$; some sets of ${\cal P}$ may be empty).
Let $G_S$  be the underlying undirected graph of $D-A_S$.
We can now use Lemma~\ref{lemY} to determine if there exist edge-disjoint forests $F_{1},F_{2},\ldots,F_{\ell}$ in $G_S$ such that
$F_i$ is $(S,P_i)$-acceptable for all $i\in [{\ell}]$ and
$V(F_i) \cap V(F_j) \subseteq S$ for all $1 \leq i < j \leq \ell$.

If such a set of forests exist then we will show that $\kappa_S(D) \geq \ell$ and if such a set of
forests do not exist for any partiton ${\cal P}$ then we will show that $\kappa_S(D) < \ell$.
By Lemma~\ref{lemY} and the fact that the  number of partitions
is bounded by $|A_S|^{\ell}$ and $|A_S| \leq 2{k \choose 2}$  would imply the desired polynomial algorithm (as
$k$ and $\ell$ are fixed).

First assume that the set of forests, $F_{1},F_{2},\ldots,F_{\ell}$, exist.
Let $H_i=\overleftrightarrow{F_i}+P_i$.
By definition of $(S,P_i)$-acceptability we observe that $H_i$ is strongly connected and $S \subseteq V(H_i)$.
Furthermore we observe that $V(H_i) \cap V(H_j) = S$ and $A(H_i) \cap A(H_j)=\emptyset$ for all $1 \leq i < j \leq \ell$ and
therefore $\kappa_S(D) \geq \ell$.

We will now show that if $\kappa_S(D) \geq \ell$ then the forests $F_{1},F_{2},\ldots,F_{\ell}$ do exist for some
partition ${\cal P}$. This will complete the proof.  Assume that $\kappa_S(D) \geq \ell$ and let
$H_1,H_2,\ldots,H_{\ell}$ be strong subgraphs in $D$ such that
$V(H_i) \cap V(H_j) = S$ and $A(H_i) \cap A(H_j)=\emptyset$ for all $1 \leq i < j \leq \ell$.
Let $P_i^*$ be the arcs from $A_S$ that belong to $H_i$.
Let $H'_i= H_i - P^*_i$ and note that $H'_i$ is symmetric.
Let $L_i$ be the undirected underlying graph of $H'_i.$ In each connected component of $L_i$ choose a spanning tree.
It remains to observe that union of the complete biorientations of the trees plus $P^*_i$ is strong since $H_i$ is strong
and each spanning tree ``preserves" connectivity of its component of $L_i$.
\end{pf}


\section{Sharp Bounds}\label{sec:bounds}

To prove the main result of this section, Theorem \ref{thm2}, we will use the two following three assertions. While the first is obvious, the second is simple, but the third is
quite a non-trivial result.

\begin{obs}\label{thm1}
If $D'$ is a strong spanning digraph of a strong digraph $D$, then
$\kappa_k(D')\leq \kappa_k(D)$.
\end{obs}

\begin{lem} \label{lemX}
For all digraphs $D$ and $k \geq 2$ we have $\kappa_k(D) \leq \delta^+(D)$ and $\kappa_k(D) \leq \delta^-(D)$.
\end{lem}
\begin{pf}
  Let $x \in V(D)$ be a vertex of minimum out-degree.
Let $S \subseteq V(D)$ be arbitrary with $|S|=k$ and let $x \in S$.
As there are $\kappa_k(D)$ arc-disjoint strong components containing $S$ and in each of these $x$ has out-degree
at least one, we must have $\delta^+(D)=d^+(x) \geq \kappa_k(D)$.
Analogously we can prove that $\kappa_k(D) \leq \delta^-(D)$.
\end{pf}

\begin{thm}(Tillson's decomposition theorem)\cite{Tillson}\label{thm01}
The arcs of $\overleftrightarrow{K}_n$ can be decomposed into
Hamiltonian cycles if and only if $n\neq 4,6$.
\end{thm}

The following result concerning the exact values of
$\kappa_k(\overleftrightarrow{K}_n)$ will be used in the proof of
the main result of this section.

\begin{lem}\label{thm3} For $2\leq k\leq n$, we have
\[
\kappa_k(\overleftrightarrow{K}_n)=\left\{
   \begin{array}{ll}
      {n-1}, & \mbox{if $2\leq k\leq n$~and~$k\not\in \{4,6\}$;}\\
     {n- 2}, &\mbox{otherwise.}
   \end{array}
   \right.
\]
\end{lem}
\begin{pf}
We first consider the case of $2\leq k=n$. By Theorem~\ref{thm01},
we clearly have $\kappa_n(\overleftrightarrow{K}_n)\geq n-1$ for
$n\notin \{4,6\}$.
Furthermore, by Lemma~\ref{lemX}, we have $\kappa_n(\overleftrightarrow{K}_n)\leq
\delta^+(\overleftrightarrow{K}_n) = n-1$ so $\kappa_n(\overleftrightarrow{K}_n)= n-1$ for $n\notin \{4,6\}$.
For $n=4$, since $K_n$ contains a Hamiltonian cycle, the two orientations of the cycle imply that
$\kappa_n(\overleftrightarrow{K}_n) \geq 2 = n-2$. To see
that there are at most two arc-disjoint strong spanning
subgraphs of $\overleftrightarrow{K}_n$, suppose that
there are three arc-disjoint such subgraphs. Then each such subgraph
must have exactly four arcs (as $|A(\overleftrightarrow{K}_n)|=12$),
and so all of these three subgraphs are
Hamiltonian cycles, which means that the arcs of
$\overleftrightarrow{K}_n$ can be decomposed into Hamiltonian
cycles, a contradiction to Theorem~\ref{thm01}). Hence,
$\kappa_n(\overleftrightarrow{K}_n)= n-2$ for $n=4$. Similarly, we
can prove that $\kappa_n(\overleftrightarrow{K}_n)= n-2$ for
$n=6$, as $K_n$ contains two edge-disjoint Hamiltonian cycles, and therefore $\overleftrightarrow{K}_n$
contains four arc-disjoint Hamiltonian cycles.

We next consider the case of $2\leq k\leq n-1$. Let $S=\{u_i\mid
1\leq i\leq k\}$ and $V(\overleftrightarrow{K}_n)\setminus
S=\{v_j\mid 1\leq j\leq n-k\}$. Let $A$ be a maximum-size set of
internally disjoint strong subgraphs containing $S$ in
$\overleftrightarrow{K}_n$. Let $A_1$ be the set of strong subgraphs
whose vertex set is $S$ and let $A_2$ be the set of strong subgraphs
in $A$ for which $S$ is a proper subset of the vertex set of each of
such strong subgraph. Hence, $A=A_1\cup A_2$. Since every strong
subgraph in $A_2$ contains at least one vertex belonging to
$V(\overleftrightarrow{K}_n)\setminus S$, we have $|A_2|\leq
|V(\overleftrightarrow{K}_n)\setminus S|=n-k$ and furthermore,
$|A_1|\leq \lfloor \frac{2{k\choose 2}}{k}\rfloor=k-1$ since each
strong subgraph containing $S$ must have at least $k$ arcs. Hence,
$|A|=|A_1|+|A_2| \leq n-1$ and so
$\kappa_k(\overleftrightarrow{K}_n)\leq
\kappa_{S}(\overleftrightarrow{K}_n)=|A|\leq n-1$ for $2\leq k\leq
n-1$. In fact, for the case of $k\in \{4,6\}$, by the argument of
the first paragraph, we have $|A_1|\leq k-2$, and so
$\kappa_k(\overleftrightarrow{K}_n)\leq n-2$ for $k\in \{4,6\}$.

If $k\notin \{4,6\}$, then in $D[S]$, there are $k-1$ edge-disjoint
Hamiltonian cycles by Theorem~\ref{thm01}. For $1\leq j\leq n-k$,
let $G_j$ be a strong subgraph with vertex set $V(G_j)=\{u_i,v_j\mid
1\leq i\leq k\}$ and arc set $A(G_j)=\{u_iv_j, v_ju_i\mid 1\leq
i\leq k\}$. So there are at least $n-1$ internally disjoint strong
subgraphs containing $S$ in $\overleftrightarrow{K}_n$, and then
$\kappa_k(\overleftrightarrow{K}_n)\geq n-1$. Hence,
$\kappa_k(\overleftrightarrow{K}_n)= n-1$ for $k\notin \{4,6\}$ and
$2\leq k\leq n-1$.

Otherwise, we have $k\in \{4,6\}$. With a similar argument, we can
obtain $n-2$ internally disjoint strong subgraphs containing $S$ in
$\overleftrightarrow{K}_n$. Hence,
$\kappa_k(\overleftrightarrow{K}_n)= n-2$ for $k\in \{4,6\}$ and
$2\leq k\leq n-1$. The concludes our proof.
\end{pf}

We now obtain a sharp lower bound and a sharp upper bound of
$\kappa_k(D)$ for $2\leq k\leq n$.

\begin{thm}\label{thm2}
Let $2\leq k\leq n$. For a strong digraph $D$ of order $n$, we have
$$1\leq \kappa_k(D)\leq n-1.$$ Moreover, both bounds are sharp, and
the upper bound holds if and only if $D\cong \overleftrightarrow{K}_n$, $2\leq k\leq n$ and $k\not\in \{4,6\}$.
\end{thm}
\begin{pf}
The lower bound is clear by the definition of $\kappa_k(D)$, and for
the sharpness, a cycle is our desired digraph. The upper bound and its
sharpness hold by Observation~\ref{thm1} and Lemma~\ref{thm3}.

If $D$ is not equal to $\overleftrightarrow{K}_n$ then $\delta^+(D) \leq n-2$ and by
Lemma~\ref{lemX} we note that $\kappa_k(D) \leq \delta^+(D) \leq n-2$. Therefore, by Lemma~\ref{thm3},
the upper bound holds if and only if $D\cong \overleftrightarrow{K}_n$, $2\leq k\leq n$ and $k\not\in \{4,6\}$.
\end{pf}

\section{Open Problems}\label{sec:openq}

We have obtained certain complexity results, in particular, showing that strong subgraph $k$-connectivity is, in a sense, harder to compute than generalized (undirected) $k$-connectivity. Several
interesting open questions remain. We conjecture that it is NP-complete to decide for fixed integers $k\ge 2$ and $\ell\ge 2$ and a given digraph $D$ whether $\kappa_k(D)\ge \ell$.
Recall that the same question is open for undirected graphs, too. We believe that further non-trivial polynomial algorithms can be obtained for computing strong subgraph $k$-connectivity in certain classes of digraphs. The {\sc Directed $k$-Linkage} problem is polynomial-time solvable for planar digraphs \cite{Schr} and digraphs of bounded directed treewidth \cite{JRST}. However, we cannot use our approach in proving Theorem \ref{thm5} directly as the structure of minimum-size strong subgraphs in these two classes of digraphs is more complicated than in semicomplete digraphs. Certainly, we cannot exclude the possibility that computing strong subgraph $k$-connectivity in planar digraphs and/or in digraphs of bounded directed treewidth is NP-complete.


\vskip 1cm
\noindent {\bf Acknowledgements.} Yuefang Sun was supported
by National Natural Science Foundation of China (No.11401389) and China Scholarship Council (No.201608330111).
Gregory Gutin was partially supported by Royal Society Wolfson Research Merit Award. Xiaoyan Zhang was supported
by National Natural Science Foundation of China (No.11471003) and Qing Lan Project.
The authors are very grateful to J{\o}ergen Bang-Jensen, Alex Scott and Magnus Wahlstr{\"o}m for useful discussions.

\end{document}